\documentclass[prl,twocolumn,showpacs,superscriptaddress]{revtex4}
\usepackage{dcolumn}
\usepackage{bm}
\usepackage{graphicx}
\usepackage{amsmath}
\usepackage{amssymb}
\usepackage{bm}
\begin{document}
\title{Magnetotransport in the CeIrIn${_5}$ system: The influence
of antiferromagnetic fluctuations}
\author{Sunil Nair}
\affiliation{Max Planck Institute for Chemical Physics of Solids,
Noethnitzer Str. 40, 01187 Dresden, Germany}
\author{M.~Nicklas}
\affiliation{Max Planck Institute for Chemical Physics of Solids,
Noethnitzer Str. 40, 01187 Dresden, Germany}
\author{J.~L.~Sarrao}
\affiliation{Los Alamos National Laboratory, Los Alamos, New
Mexico 87545, USA}
\author{J.~D.~Thompson}
\affiliation{Los Alamos National Laboratory, Los Alamos, New
Mexico 87545, USA}
\author{F.~Steglich}
\affiliation{Max Planck Institute for Chemical Physics of Solids,
Noethnitzer Str. 40, 01187 Dresden, Germany}
\author{S.~Wirth}
\affiliation{Max Planck Institute for Chemical Physics of Solids,
Noethnitzer Str. 40, 01187 Dresden, Germany}
\date{\today}
\begin{abstract}
We present an overview of magnetotransport measurements on the
heavy-fermion superconductor CeIrIn$_5$. Sensitive measurements of
the Hall effect and magnetoresistance (MR) are used to elucidate
the low temperature phase diagram of this system. The normal-state
magnetotransport is highly anomalous, and experimental signatures
of a pseudogap-like precursor state to superconductivity as well
as evidence for two distinct scattering times governing the Hall
effect and the MR are observed. Our observations point out the
influence of antiferromagnetic fluctuations on the
magnetotransport in this class of materials. The implications of
these findings, both in the context of unconventional
superconductivity in heavy-fermion systems as well as in relation
to the high temperature superconducting cuprates are discussed.
\end{abstract}
\pacs{Heavy Fermion superconductors, Hall effect,
Magnetoresistance} \maketitle

\hspace*{-0.3cm}{\bf I. INTRODUCTION}\\[0.3cm]
Superconductivity---the zero electrical resistance state of
matter---involves the formation of bound pairs of electrons,
so-called Cooper Pairs. The BCS theory \cite{bar} states that
these electrons are bound together by dynamic lattice distortions
(or phonons). This theory provides a remarkably accurate
description for superconductivity in many, if not all,
``conventional'' superconductors. Among the so-called
``unconventional'' superconductors, in particular two contrasting
classes of materials---the heavy-fermion and the high-temperature
cuprate superconductors---have posed an irrefutable challenge to
this theory.  The non-compliance with the standard BCS model is
primarily based on the failure to reconcile how magnetism could
coexist with the superconducting ground state in these systems. In
conventional superconductors, it is known that tiny amounts of
magnetic impurties rapidly destabilize the superconducting
condensate. The aforementioned two families of superconductors on
the other hand, are comprised of dense arrays of magnetic ions.
The observation of superconductivity in an inherently magnetic
environment \cite{ste} has led researchers to postulate that
(antiferro-)magnetic fluctuations themselves were the bosonic mode
coupling the electrons into Cooper pairs, thus facilitating a
superconducting ground state in these materials. The observation
of superconductivity in the vicinity of antiferromagnetic quantum
critical points (QCP, zero-temperature continuous phase
transitions) in some heavy-fermion systems \cite{mathur} has lent
further credence to this theory.

The contrast between these two families of materials is stark.
Metals are typically well described by the Landau Fermi liquid
(LFL) theory. Although the heavy-fermion systems exhibit metallic
behavior, this theory cannot be applied in certain cases, e.g. in
the vicinity of a magnetic QCP that gives rise to an anomalous
metallic state \cite{sch}. The parent high temperature
superconductors on the other hand are Mott insulators: systems
where strong electron-electron correlations promote an insulating
ground state at the cost of a conducting one that would be
expected from conventional band theories. In spite of this
apparent difference, a surprising similarity in both the
superconducting and normal-state properties of some heavy-fermion
systems with the high-$T_c$ cuprates has been observed. This has
been brought to attention by recent observations in the
Ce$M$In$_5$ ($M$ = Co, Ir, Rh) family of compounds \cite{sar}.
These quasi-two-dimensional systems have revealed a host of
experimental signatures which are remarkably similar to those
observed earlier in the high-$T_c$ cuprates \cite{nak}. For
instance, the normal-state resistivity is seen to be linear as a
function of temperature, the Hall effect is strongly temperature
dependent, and an anomalous Nernst effect is observed above the
superconducting upper critical field $H_{c2}$ \cite{bel}. The
superconducting state is also unconventional, and it was suggested
that the superconducting gap possibly has line nodes consistent
with $d$-wave symmetry \cite{mat}. Here we summarize some of our
recent results on the normal-state magnetotransport in the
compound CeIrIn$_5$. Using sensitive measurements of the Hall
effect and magnetoresistance (MR), a comprehensive low-temperature
phase diagram of this system is charted out. The implications of
our observations are discussed, both in terms of discerning the
influence of antiferromagnetic fluctuations on the
magnetotransport, as well as in placing these heavy-fermion
superconductors in perspective to the superconducting cuprates.
\\[0.2cm]

\hspace*{-0.3cm}{\bf II. EXPERIMENTAL TECHNIQUES}\\[0.3cm]
Measurements of the Hall effect and MR are conducted on a single
crystalline sample of CeIrIn$_5$ within the temperature range 0.05
K $\le T \le$ 2.5 K in the form of isothermal field sweeps. The
magnetic field (of up to 15 T) is applied parallel to the
crystallographic $c$ axis, and the Hall voltage is extracted as
the asymmetric component under magnetic field reversal.
Low-temperature transformers are used in conjunction with
low-noise voltage pre-amplifiers to enable an effective resolution
of better than $\pm$0.01 nV.\\[0.2cm]

\hspace*{-0.3cm}{\bf III. RESULTS AND DISCUSSION}\\[0.3cm]
The system CeIrIn$_5$, which is reported to have a resistive
transition temperature of 1.2 K \cite{pet} is probably the most
enigmatic one in the CeMIn$_5$ series of compounds: the
superconductivity in this system seems to be separated from the
magnetic instability in the field-temperature-pressure phase
space. In the other primary members of this family (CeCoIn$_5$ and
CeRhIn$_5$) it is known that superconductivity is closely tied to
the suppression of magnetic order. For instance, in the
ambient-pressure superconductor CeCoIn$_5$, the magnetic
instability is proposed to be in the vicinity of $H_{c2}$
\cite{pag}. However, in CeIrIn$_5$, the magnetic instability
(which was suggested to be metamagnetic in origin) is speculated
to lie above 25 T, whereas $H_{c2}$ is only of the order of 3 T
\cite{cap}. Though the range of our measurements does not
encompass the magnetic instability, this contrast is also clearly
manifested in our data. In systems exhibiting a typical
antiferromagnetic QCP (believed to be the case in CeCoIn$_5$), an
applied magnetic field destabilizes the antiferromagnetic order
and facilitates the recovery of the LFL regime. In CeIrIn$_5$,
however, a reverse trend is observed wherein the applied magnetic
field suppresses (and not enhances) the LFL region. Thus the bulk
superconductivity in CeIrIn$_5$ appears to emerge from {\em
within} the LFL region.

\begin{figure}
\includegraphics[width=7.4cm,clip]{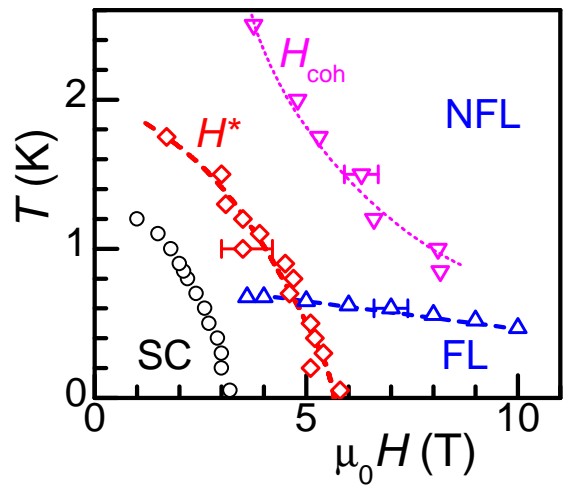}
\caption{$H$--$T$ phase diagram of CeIrIn$_5$ as determined from a
combination of Hall-effect and MR measurements.} \label{fig1}
\end{figure}

Fig.~\ref{fig1} summarizes the phase diagram of CeIrIn$_5$ as
determined from our magnetotransport measurements \cite{nai}. Two
distinguishable crossover lines in this phase diagram are clearly
related to the magnetic instability at about 25 T: (i) The LFL to
non-Fermi liquid (NFL) crossover, as determined by deviations from
Kohler's scaling rule and (ii) the onset of a coherent-Kondo
scattering regime, as determined from a crossover in the sign of
the MR at $H_{coh}$.  The LFL--NFL crossover is seen to be in good
agreement with prior reports, and a linear extrapolation indeed
intercepts the magnetic field axis at about 25 T. Though the
functional form of the coherent to incoherent Kondo regime is
non-trivial, a crossing between these two lines is unlikely, and
the Kondo coherence would also be expected to vanish at the
magnetic instability.

A striking result from the analysis of our magnetotransport data
pertains to the observation of a precursor state to
superconductivity in CeIrIn$_5$. This state which is seen as a
curve that envelops $H_{c2}$ in Fig.~\ref{fig1} was inferred from
the field-dependence of the Hall angle $\theta_H$. Though rarely
utilized as a means of investigating heavy-fermion systems, prior
work in the superconducting cuprates has demonstrated that its
cotangent $\cot \theta_H = \rho_{xx} / \rho_{xy}$ is a quantity of
fundamental interest \cite{chi}. Besides being a measure of the
charge carrier mobility, a quadratic temperature dependence of
$\cot \theta_H$ was observed in the cuprates independent of the
charge carrier density and the extent of impurity substitution.
Since the resistivity $\rho_{xx}$ is linear in $T$, this $T^2$
dependence of $\cot \theta_H$ is thought to be a manifestation of
the fact that there are two distinct scattering rates which
independently influence the resistivity and the Hall effect
($\tau_{\rm tr}$ and $\tau_H$, respectively). Moreover, deviations
from the $T^2$ behavior of $\cot \theta_H$ were interpreted in
terms of the onset of the pseudogap state in the cuprates
\cite{abe}. Our measurement protocol (we conducted isothermal
field sweeps) enables us to investigate the magnetic-field
dependence of this quantity in detail, and $\cot \theta_H$ is seen
to have an $H^{-1}$ dependence within a substantial region of the
$H$--$T$ phase space. In the vicinity of the superconducting
region systematic deviations from this $H^{-1}$ dependence are
seen (starting at a critical field $H^*$), the field and
temperature dependence of which is exhibited in the
Fig.~\ref{fig1}. Our measurements clearly indicate the existence
of a pseudogap-like precursor state to superconductivity in
CeIrIn$_5$, a phenomenon which may be generic to many other
heavy-fermion superconductors as well. Interestingly, the critical
field of this precursor state scales onto the superconducting
upper critical field $H_{c2}$, implying that both of them might
arise from the same underlying physical mechanism \cite{nai}.

\begin{figure}
\includegraphics[width=8.0cm,clip]{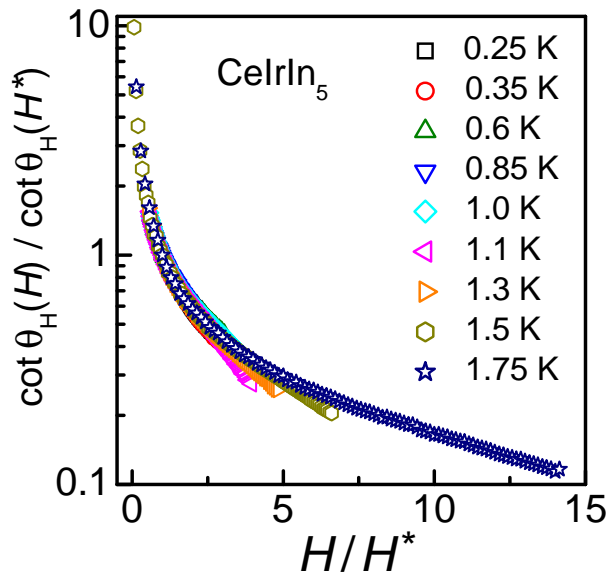}
\caption{Scaling of $\cot \theta_H$ and the field $H$ normalized
with respect to the values at $H^*(T)$ which demarcates the
precursor state. All the isotherms are seen to collapse on a
single generic curve.} \label{fig2}
\end{figure}

The presence of this precursor state appears to crucially
influence the normal-state magnetotransport in CeIrIn$_5$.
Moreover, it seems to influence $\tau_{\rm tr}$ and $\tau_H$ in a
disparate fashion. This is clearly borne out by two key
experimental observations: Firstly, contrary to prior reports
\cite{nak2} the modified Kohler's scaling (which relates the MR to
the Hall angle) breaks down in the precursor state. Secondly, a
model-dependent single parameter scaling of magnetotransport
quantities---using the demarcation $H^*(T)$ of the precursor
state---is seen to be applicable {\em only} for the Hall angle
\cite{nai2}. This is shown in Fig.~\ref{fig2}, where the
normalized Hall angle $\cot \theta_H(H) / \cot \theta_H(H^*)$ is
plotted as a function of the normalized field ($H / H^*$)
revealing good scaling behavior. The fact that neither the
resistivity, nor the Hall effect individually exhibit this scaling
implies that the precursor state preferentially influences the
Hall channel. This scaling clearly demonstrates that the precursor
state in CeIrIn$_5$ represents a fundamental energy scale of the
system; in addition to the well recognized energy scales
corresponding to the crystal electric field, the intersite
coupling and the single-ion Kondo effect. It also re-emphasizes
the presence of two distinct scattering times in these systems, in
similarity to observations in the superconducting cuprates. Though
the existence of two scattering times in heavy-fermion systems
remains to be investigated in adequate detail, this scenario might
well be applicable to other systems.

Our key observations, namely an $H^{-1}$ dependence of the Hall
angle (which in turn is used to demarcate the precursor state) and
two distinct scattering times can be adequately described by two
theories which have been extensively used earlier in the cuprates.
The spin-charge-separation scenario developed by Anderson
\cite{and} postulates the formation of two different
quasiparticles each of which is associated with the spin and
charge degrees of freedom. The two different scattering times then
correspond to dissimilar scattering events associated with these
two different kinds of quasiparticles. On the other hand, the
nearly antiferromagnetic Fermi liquid (NAFL) scenario \cite{sto}
postulates the modification of scattering rates along different
regions of the Fermi surface. This is accomplished by the
formation of so-called {\em hot spots} which represent regions of
the Fermi surface where it intersects the antiferromagnetic
Brillioun zone and where scattering becomes singular. Thus, all
the transport coefficients are renormalized with respect to the
anisotropy associated with the different regions of the Fermi
surface. Interestingly, both of these scenarios can explain the
observed $H^{-1}$ dependence of the Hall angle.

While the spin-charge-separation scenario cannot be ruled out, the
NAFL picture is a particularly attractive one for the Ce$M$In$_5$
compounds. In CeCoIn$_5$ for instance, an anisotropic destruction
of the Fermi surface in the limit $T \! \rightarrow \! 0$ was
inferred, analogous to the pseudogap phase in the cuprates
\cite{tan}. It was suggested that this destruction of the Fermi
surface could be caused by uniaxial spin fluctuations. Though the
superconductivity in CeIrIn$_5$ is reasonably separated from the
magnetic instability in the $H$--$T$ phase space, the presence of
antiferromagnetic fluctuations in the regime investigated here has
also been inferred from Nuclear Quadrupole Resonance (NQR)
measurements \cite{zhe}. These fluctuations themselves are thought
to be anisotropic, with the magnetic correlation length along the
basal plane being substantially larger than the one along the $c$
axis. Thermal conductivity measurements have suggested that the
superconducting gap in this system may have a $d_{x^2 - y^2}$
symmetry, which is consistent with a scenario in which
superconductivity is mediated by incipient antiferromagnetic
fluctuations \cite{kas}. The fact that antiferromagnetic spin
fluctuations are a crucial ingredient in determining the
electronic ground state of these systems was also demonstrated by
a chemical substitution study \cite{pha}: small amounts of Cd
substitution on the In site shift the delicate balance between
magnetism and superconductivity. The evolution of a magnetically
ordered ground state with Cd substitution could not be accounted
for by simple unit-cell volume considerations implying that the
observed effects were a consequence of tailoring the electronic
band structure.

A phenomenon in the cuprate superconductors which had no apparent
analogue in the heavy-fermion systems was the formation of charge
stripes. Such stripes would allow for both, antiferromagnetism
(with localized spins) and superconductivity (with mobile charge
carriers) to coexist within the Cu-O plane, by promoting spatial
segregation of holes into stripe-like patterns. Although evidence
for the existence of such charge stripes have been obtained
\cite{tra} it remains to be clarified whether these stripes would
be conducive (or detrimental) to the formation of a
superconducting ground state. Interestingly, it has recently been
predicted theoretically that the existence of a striped phase in
heavy-fermion systems is also a distinct possibility \cite{zhu}.
It was suggested that these so-called Kondo-stripes separate
electronically inhomogenous heavy Fermi liquid phases and result
in a modulation of the charge density, an effect which should be
visible experimentally. If verified, it would be interesting to
see the manner in which heavy-fermion superconductivity is
influenced by the existence of such charge density modulations.

In summary, we have discussed the implications of our
magnetotransport data obtained on CeIrIn$_5$ and their relation to
prior observations on the high-temperature superconducting
cuprates. Two of our key observations---a pseudogap-like precursor
state and the existence of two distinct scattering times---are
reminiscent of the behavior in the cuprates. These observations
are also consistent with a scenario in which incipient
antiferromagnetic fluctuations crucially influence the
magnetotransport in these two disparate classes of systems by
modifying the electron scattering rates along different parts of
the Fermi surface. The phenomenon of unconventional
superconductivity in the heavy-fermion systems, the high-$T_c$
cuprates, and even the newly discovered iron-oxypnictides can thus
possibly be placed on a universal platform: the nature and
properties of the electronic ground state are dictated by magnetic
fluctuations.\\[0.2cm]

\hspace*{-0.3cm}{\bf IV. ACKNOWLEDGEMENTS}\\[0.3cm]
S.N. was supported by the Alexander von Humboldt Foundation. Work
at Dresden was supported by the EC (CoMePhS 517039) and the DFG
(Forschergruppe 960). Work at Los Alamos was performed under the
auspices of the U.S. Department of Energy/Office of Science.

\end{document}